\documentclass[10pt, conference, compsocconf]{IEEEtran}
\usepackage{cite}
\usepackage{mathtools}
\usepackage{gensymb}
\usepackage{tabu}
\usepackage[bf]{caption}
\usepackage{amsmath}
\usepackage{algorithm}
\usepackage[noend]{algpseudocode}
\usepackage{breqn}
\usepackage{subcaption}
\usepackage{caption}
\usepackage{float}
\usepackage{float}

\begin{document}
\title{DSIC: Deep Learning based Self-Interference Cancellation for In-Band Full Duplex Wireless}

\author
{\IEEEauthorblockN{Hanqing Guo, Nan Zhang, Saeed AlQarni, Shaoen Wu} 
\IEEEauthorblockA{Ball State University\\
Muncie, IN USA \\
\{hguo, nzhang, saalqarni, swu\}@bsu.edu
}
}

\maketitle

\begin{abstract}
In-band full duplex wireless is of utmost interest to future wireless communication and networking due to great potentials of spectrum efficiency. IBFD wireless, however, is throttled by its key challenge, namely self-interference. Therefore, effective self-interference cancellation is the key to enable IBFD wireless. This paper proposes a real-time non-linear self-interference cancellation solution based on deep learning. In this solution, a self-interference channel is modeled by a deep neural network (DNN).  Synchronized self-interference channel data is first collected to train the DNN of the self-interference channel. Afterwards, the trained DNN is used to cancel the self-interference at a wireless node. This solution has been implemented on a USRP SDR testbed and evaluated in real world in multiple scenarios with various modulations in transmitting information including numbers, texts as well as images. It results in the performance of 17dB in digital cancellation, which is very close to the self-interference power and nearly cancels the self-interference at a SDR node in the testbed.  The solution yields an average of 8.5\% bit error rate (BER) over many scenarios and different modulation schemes.

\end{abstract}

%
\IEEEpeerreviewmaketitle

\section{Introduction}
\label{sec: intro}
Traditional wireless communication, widely used today, supports bi-directional transmissions in the way of time-division duplex (TDD) or frequency-division duplex (FDD). In-band full duplex (IBFD) wireless refers to the bi-directional transmission on the same frequency band at the same time, which intuitively boosts the spectrum efficiency and channel utilization. As the demand of wireless transmission is sharply increasing, IBFD wireless is very promising to future wireless communications. 

\begin{figure}[h]
\centering
\includegraphics[width=.5\textwidth]{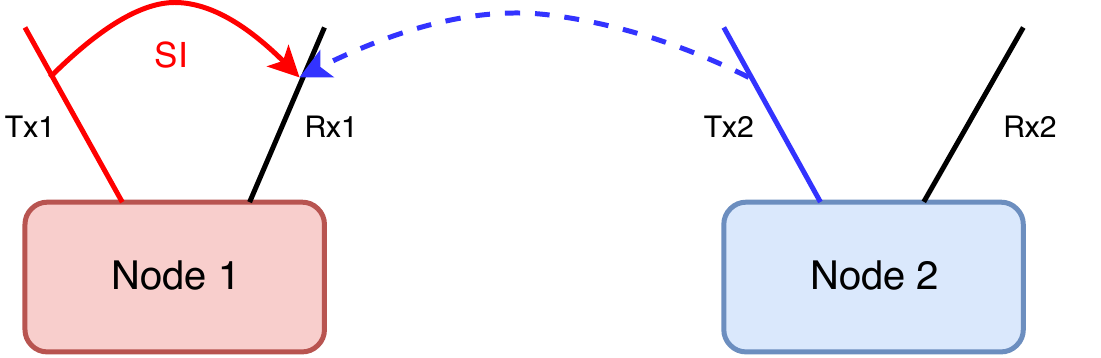}
\vspace{-0.1in}
\caption{IBFD Self-Interference Scenario}
\vspace{-0.3in}
\label{fig:mix_input}
\end{figure}

Although IBFD wireless has been studied for years in theory and analysis, it has been progressing very slowly for practical use because of a very challenging issue called self-interference to be addressed in engineering. Refer to the IBFD wireless architecture shown in figure \ref{fig:mix_input} where the wireless node Node1 is receiving data $[101,102,103,...,120]$ via its antenna RX1 from the TX2 of Node2, while it is also transmitting data via its antenna TX1. The consequence is that the power emitted by the TX1 of Node1 will interfere the TX2's signal received at the RX1 of Node1. Such interference is called self-interference (SI), which completely overshadows the received signal and leads to the communication failure from Node2 to Node1. To enable IBFD wireless, SI has to be cancelled so that the overshadowed received signal can be recovered. Although some solutions have been proposed in the literature, their performance is not satisfactory. 

In this paper, we target self-interference cancellation (SIC), the toughest issue in IBFD wireless. We propose a non-linear SIC solution based on deep learning. We have implemented this solution on a USRP SDR testbed and evaluated its performance in the real world with this prototype. In this work, we have answered three questions: (1) how to collect synchronized wireless channel information to train the deep learning model? (2) how to model a wireless channel with a deep neural network? (3) how to implement an open source SDR IBFD wireless framework to work in real world?.
	
In the rest of this paper, Section ~\ref{sec:relate} reviews the background of IBFD wireless and literature solutions related to our work. Next, Section \ref{sec:design} presents the SI problem with a simulation, then formulates the theoretical foundation for digital SIC solutions, and describes the design of our deep learning SIC solution in details. Then, Section \ref{sec:experiments} presents an open source SDR framework implementing IBFD wireless prototypes, followed by Section \ref{sec:performance} summarizing the performance of our solution in the tests. This work is finally concluded in Section \ref{sec: conclusion}.

\section{Related Work}\label{sec:relate}
Since the first systematic study of IBFD wireless on a narrow band \cite{S.Chen1998}, several solutions have been proposed for IBFD wireless on larger bandwidths. These IBFD wireless SIC solutions can be classified into 3 different categories\cite{kim2015survey}: (1) antenna cancellation, (2) analog cancellation and (3) digital cancellation. 

In antenna cancellation, Bliss \cite{Bliss2007} showed that two transmission antennas are placed at designated positions with $\ell$ and $\ell$+$\frac{\lambda}{2}$ away from receiving antenna can cancel the SI power at its receiver, where $\lambda$ is the wave length of the carrier frequency. Other solutions used directional antennas \cite{K.Haneda2010} or MIMO antenna \cite{aryafar2012midu} or single antenna design \cite{Duarte2010, Bharadia2013}, which can cancel up to 15 dB of SI.

As for analog cancellation, a team at Rice University presented a model to dynamically adjust the gains of $I$ and $Q$ components of a signal as well as the delays to emulate an inverse  of the transmitted signal, then add the emulated signals to the receiver signal processing chain, and thus suppress the SI \cite{Duarte2010, Duarte2011}. 

Digital cancellation can be further classified into linear digital cancellation and non-linear digital cancellation. Linear digital cancellation aims to cancel the deformed digital signals passing over the environment channel, while non-linear digital cancellation targets at cancelling non-linear cubic and higher order elements generated by radio circuits \cite{Bharadia2013, Askar2014}. A team at Stanford University designed an FIR filter to emulate the components of the line of sight signals, then used at modeled \textit{SI} channel to emulate and generate  the SI signals at a full duplex node \cite{Bharadia2013}. They also included a non-linear component in SIC by using a general model in the form of Taylor Series Expansion.  Another work \cite{sim2017nonlinear} proposed a pre-calibration-based cancellation technique that linearizes a transmitter and cancels SI with linear-only cancellation at a receiver, which basically, converted a non-linear problem to a linear problem and then solved it.

As far as the performance of IBFD wireless solutions is concerned, some works have reported that the channel capacity could be doubled on a single-hop link, while in mesh network the actual benefit of IBFD wireless may be effected by spatial reuse and asynchronous contention \cite{xie2014does}. Another works \cite{Bharadia2013, kim2013performance} shows 1.87 times throughput improvement and 1.42 times capacity over a half-duplex system at a transmission power of 20 dBm. A recent work \cite{al2017performance} shows that at a certain situation, IBFD wireless can double the downlink and uplink throughputs of static TDD. Another recent work reported a performance improvement by 30\% to 66\% compared to the 5G Dynamic TDD \cite{al2017performance}.

\section{Deep Learning Based Digital Self-Interference Cancellation} \label{sec:design}
\subsection{SI Problem Formulation} \label{subsec: problem}
Revisit Figure \ref{fig:mix_input} where there are two full-duplex devices Node1 and Node2, and both of them can transmit (TX) and receive (RX) at same time and on the same channel. Suppose that Node1 wants to send a vector stream of $[1,2,3,...,10]$ to Node2, while Node2 sends a stream of $[101,102,103,...,110]$ to Node1. How can we achieve this without dividing the channel into uplink and downlink two wireless sub-channels? Since RX1 will get the overwhelming power from TX1 as SI, while the power from TX2 is expected far less than \textit{SI} assuming that both devices are using the same transmission level,  what does the signal mixture of the SI and the received signal from TX2 will look like? 

A simple test was designed to illustrate this problem vividly. We set up two USRP nodes as in Figure \ref{fig:mix_input}, one with both transmission (TX1) and receiving (RX1), and the other one only for transmission (TX2). They were placed one meter apart from each other. Both of them use the channel of 2.5GHz, while TX1 transmits the sequence $[1,2,...,20]$, and TX2 transmits the sequence $[101,102,...,120]$. 

\begin{figure}[h]
\vspace{-0.15in}
\centering
\includegraphics[width=.45\textwidth]{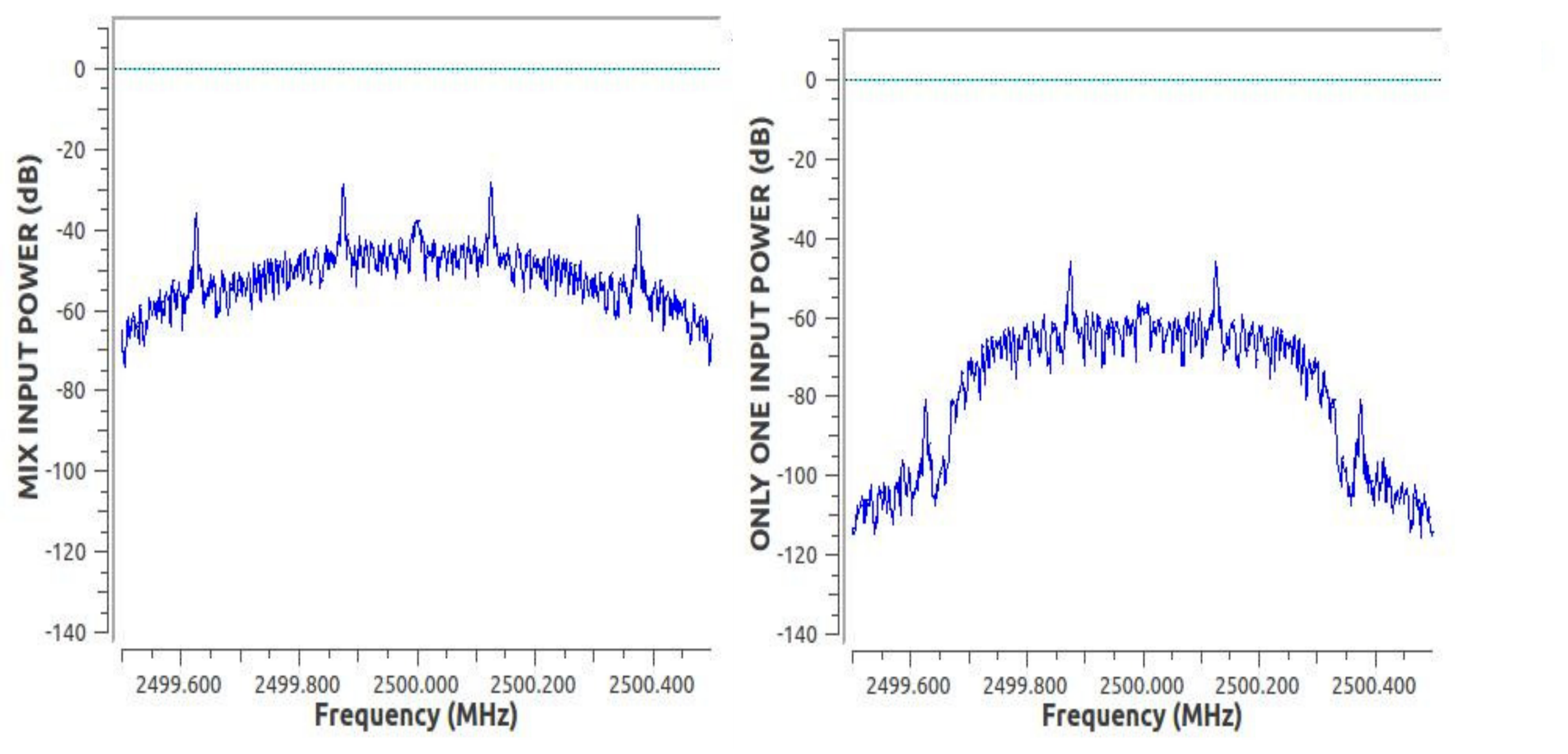}
\vspace{-0.05in}
\caption{Compare with mix input and single input}
\vspace{-0.2in}
\label{fig:power_mix_input}
\end{figure}

The left plot and the fight plot of Figure \ref{fig:power_mix_input} respectively show the received signal powers at RX1 when TX2 is enabled and disabled. When only TX2 sends signals,  the relative received signal power at RX1 is around $-40$ dBm as shown on the right plot on the figure (\ref{fig:power_mix_input}). When TX1 signals are enabled in addition to the signals of TX2, the relative received signal power of RX1 is gained to $-20$dBm as shown on the left plot, which means the signals on the same frequency from TX1 and TX2 mix together and result in the much higher received power at RX1. The interference problem of SI is clearly disasterous from this measurement. 
\vspace{-0.2in}
\subsection{Theoretical Foundation of Digital Self-Interference Cancellation}\label{sec:solution}
To cancel the SI as shown in above section and recover the received useful signal from others, we propose a deep learning based digital SI cancellation. The basic concept of the digital SI cancellation is simply. Suppose that, at a wireless node, the self-transmitted signal $x[n]$ passes through the SI channel and is transformed into $y_{si}[n]$ at the receiver, and denote $y[n]$ the received signal at the receiver that is a mixture of the SI signal $y_{si}[n]$ and the useful signal $m[n]$ from another node. The following formula (\ref{equ:novel_recovery}) can be used to recover the useful signal $m[n]$. Since the system knows of its self-transmitted signal $x[n]$ to the SI channel and the received mixture signal $y[n]$, the key challenge to recover the useful signal $m[n]$ is to compute the SI $y_{si}[n]$ based on the known signal $x[n]$.
\begin{equation}
m[n] = y[n] - y_{si}[n]
\label{equ:novel_recovery}
\end{equation}

One observation is that the self-transmitted signal $x[n]$ can be decomposed as the sum of different impulse signals $\delta$ as below:
\begin{equation}
x[n]=\displaystyle\sum_{k=0}^K x[k]\delta[n-k]
\label{equ:decompose}
\end{equation} 
where $\delta[n]$ is:
\begin{equation}
  \delta[n]=\begin{cases}
    1, & \text{if $n = 0$}.\\
    0, & \text{otherwise}.
  \end{cases}
\label{equ: impulse}
\end{equation}

Denote  $h[n]$ the impulse response of SI channel, then the SI signal $y_{si}[n]$ can be calculated as:
\begin{equation}
\begin{split}
y_{si}[n] &= \displaystyle\sum_{k=0}^K x[k]h[n-k]\\
	 &= \displaystyle\sum_{k=0}^K h[k]x[n-k]
\end{split}		 
\label{equ:y_si}
\end{equation} 
Based on Formulas (\ref{equ:novel_recovery}) and (\ref{equ:y_si}), the useful message $m[n]$ can be obtained if we know $h[n]$. In practice, however, we can't obtain any precise SI channel impulse response $h[n]$ beforehand. Therefore, what we can do is to estimate the SI channel response $\hat{h}[n]$ as accurate as possible, which is used to replace the real impulse response $h[n]$ in estimating the SI signal $\hat{y_{si}}[n]$ as below where $K$ refers to the number of taps:
\begin{equation}
\hat{y_{si}}[n]=\displaystyle\sum_{k=0}^K \hat{h}[k]x[n-k]
\label{equ:eva_y_si}
\end{equation} 
\begin{equation}
\hat{m}[n] = y[n] - \hat{y}_{si}[n]
\label{equ:eva_recover}
\end{equation} 
to get $\hat{m}[n]$ as recovered useful signal.

\begin{figure*}
\centering
\includegraphics[width=\textwidth]{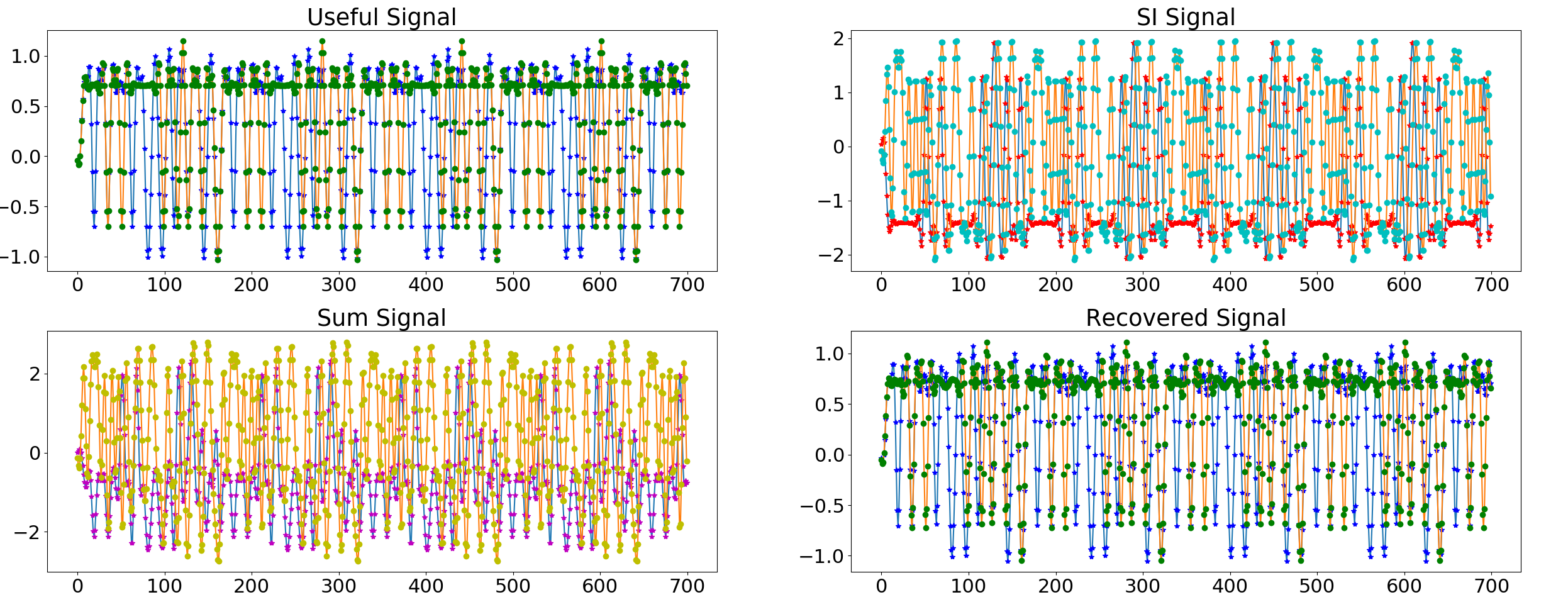}
\vspace{-0.1in}
\caption{Digital Cancellation Validation Results} 
\vspace{-0.2in}
\label{fig:recovery}
\end{figure*}

\begin{figure}[h]
\vspace{-0.1in}
\centering
\includegraphics[width=.4\textwidth]{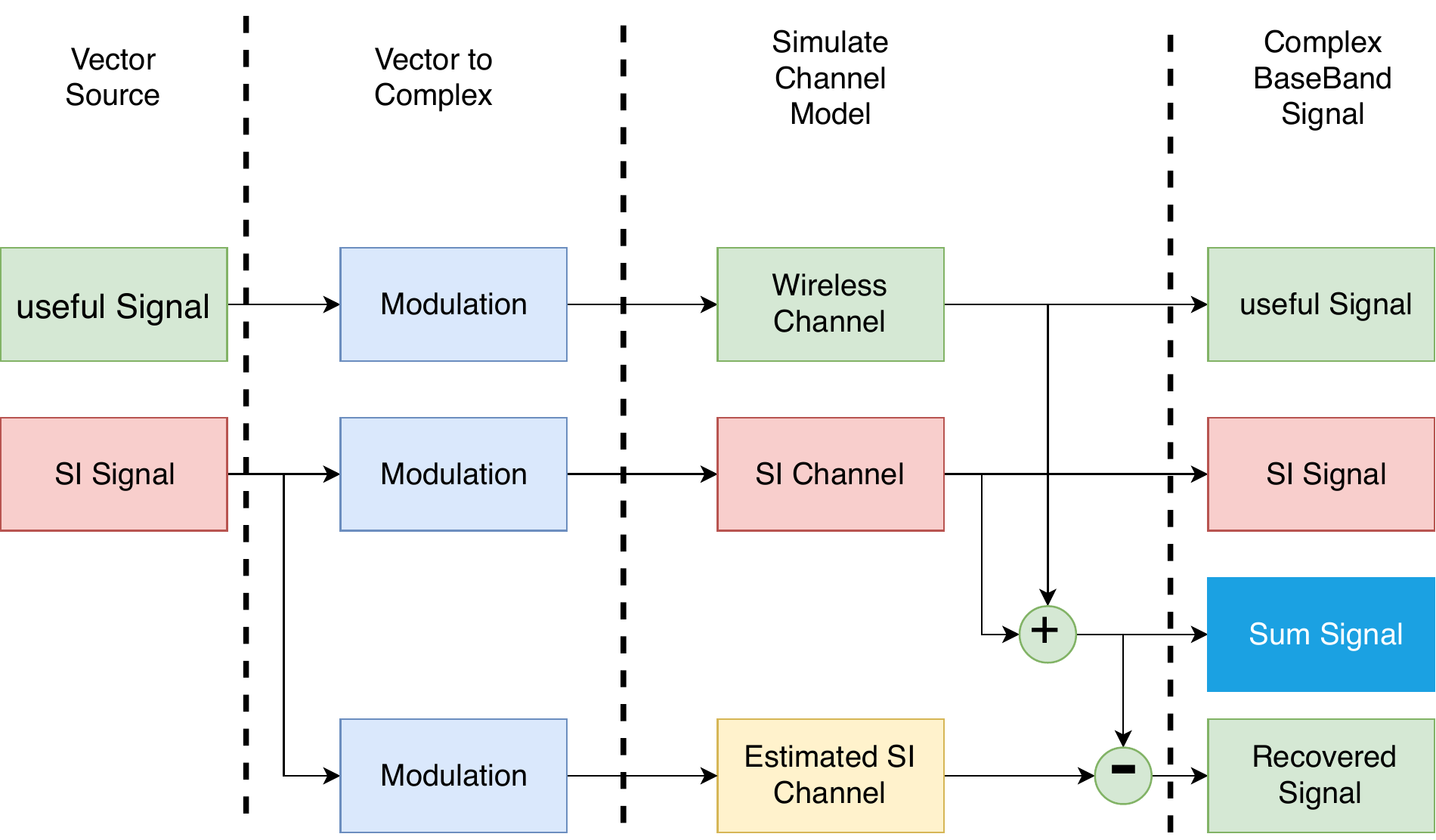}
\caption{Digital Cancellation Validation Framework}
\label{fig:recovery_principle}
\vspace{-0.1in}
\end{figure}

To demonstrate the feasibility of the digital cancellation concept, a simulation was performed on a GNU Radio SDR platform, whose framework was implemented as in Figure~\ref{fig:recovery_principle}. There are two digital vector sources: useful signal $m[n]$ being $[1,2,...,10]$, and the SI signal vector $[101,102,...,110]$, as shown on the top plots in Figure \ref{fig:recovery_principle}, Both of them use the QPSK modulation but pass through different simulated channel models of $h[k]$, which are determined by the settings of taps. The taps of the useful signal $m[n]$ {\it wireless channel} are $[1]$, while the taps in the {\it SI channel} are $[1,1]$, and the {\it estimated SI channel} taps are $[1.1,0.9]$.   After passing through channel models, all of those three streams go into Decimating FIR filters, which will be described in Section \ref{sec:experiments}). Then the first time sink receives $m[n]$, the second time sink receives $y_{si}[n]$, the third time sink receives $y[n]$, and the last time sink gets the recovered signal $\hat{m}[n]$ by using formula (\ref{equ:eva_recover}).

The simulation result shown on the bottom plots in Figure (\ref{fig:recovery}) indicates that, even though the estimated SI channel response $\hat{h}[k]$ is not completely the same as $h[k]$, the recovered signal $\hat{m}[n]$ (on the right bottom plot) is actually very similar to the useful signal $m[n]$ (one the left top plot), in that the real and image parts of those signals have very similar shapes, and the relation between the real and image numbers of signal I-Q  phases is almost kept. This result proves that if $\hat{h}[k]$ are estimated accurately, then $m[n]$ can be recovered from $y[n]$, even if $y[n]$ is severely polluted by the SI signal $y_{si}[n]$.

\subsection{Deep Learning Digital Cancellation}\label{subsec:design}
As discussed above, the core of digital SI cancellation is to estimate the SI channel response $\hat{h}[k]$. Rather than modeling and estimating the SI channel with traditional wireless channel models as most of literature solutions do, we propose to model the SI channel response with a deep neural network (DNN).  

\subsubsection{SI Channel Modeling with DNN}\label{sec:net}
\begin{figure}[h]
\includegraphics[width=.5\textwidth]{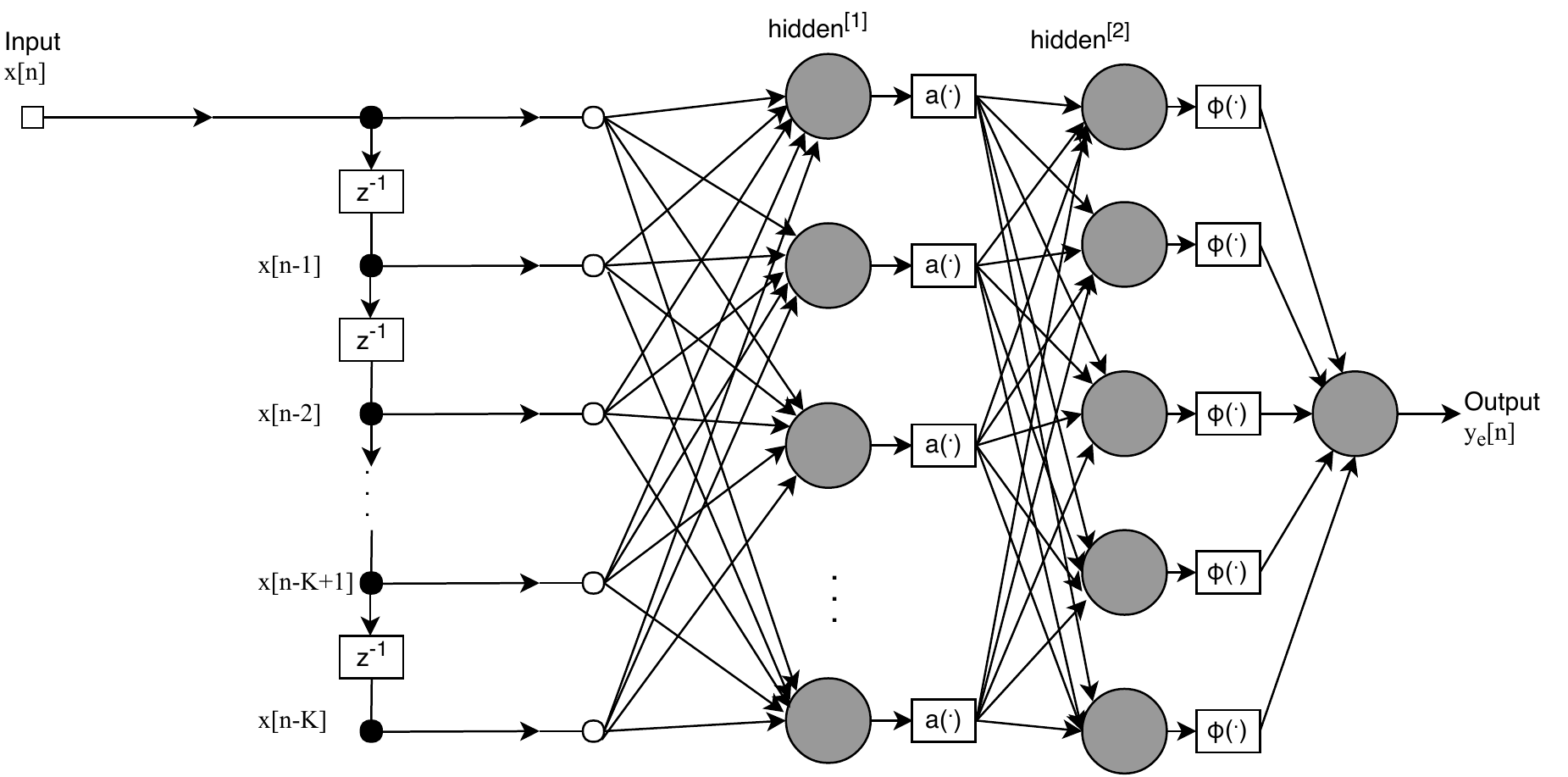}
\caption{DNN Model for SI Channel}
\vspace{-0.3in}
\label{fig:network}
\end{figure}

A DNN model is introduced to estimate the SI channel and its structure is shown in Figure (\ref{fig:network}).

\paragraph{Input and Output}
In this DNN model, the input is denoted as $x[n], x[n-1],...,x[n-K]$, and the output is referred as $y_{e}[n]$. Instead of modeling the SI channel as a one-to-one mapping, we model it as a many-to-one DNN, because the many-to-one model has a similar structure to an FIR filter. Considering Formula (\ref{equ:eva_y_si}), to estimate $\hat{y_{si}}$, we need to compute the sum of $\hat{h}[k]x[n-k]$, where $k$ varies from $0$ to $K$, which means that every output of the SI channel not only relates to the current input $x[n]$, but also previous $K$ samples of $x$. 

\paragraph{DNN Structure}
\begin{figure}[h]
\includegraphics[width=.5\textwidth]{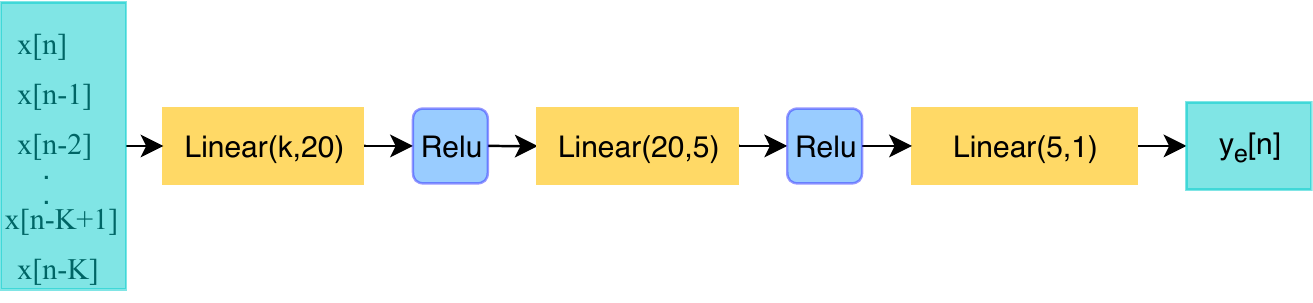}
\vspace{-0.2in}
\caption{Simplified Network}
\vspace{-0.3in}
\label{fig:simplified net}
\end{figure}

The simplified and parameterized structure of the designed DNN is shown in Figure (\ref{fig:simplified net}). From the figures of (\ref{fig:network}) and (\ref{fig:simplified net}), this DNN has 2 hidden layers, each of which is followed by an activation function $a(\cdot)$ or $\Phi(\cdot)$. The activation function enables the channel modeling a non-linear structure. We choose the Relu function for both $a(\cdot)$ and $\Phi(\cdot)$. It should be noted that there is no activation function after the last layer, because the DNN is used to compute mapping values, whose range could be any possible rational number, and it doesn't need something like Softmax or any other classification function. The first linear block takes $K$ input signals and generates an output of $1*20$ dimensions, where $K$ can be seen as the size of FIR filter taps and means how many samples are related to output sample. Then the second linear block takes $1*20$ input signals and generates an output of $1*5$ dimensions, and the last linear block takes the $1*20$ signals and output a value of $1*1$ dimension. In each linear block, the weight matrix and tge bias matrix are respectively denoted as  $W^{[l]}$ and $b^{[l]}$, where $l$ is the layer number in the DNN. Denote $o^{[l]}$ the output of linear block, the output of an activation function is referred as $a^{[l]}$. As a result, 
\vspace{-0.1in}
\[ W^{[1]} = 
\begin{bmatrix}
    w^{[1]}_{1,1} & w^{[1]}_{1,2} & w^{[1]}_{1,3} & \dots  & w^{[1]}_{1,20} \\
    w^{[1]}_{2,1} & w^{[1]}_{2,2} & w^{[1]}_{2,3} & \dots  & w^{[1]}_{2,20} \\
    \vdots & \vdots & \vdots & \ddots & \vdots \\
    w^{[1]}_{K,1} & w^{[1]}_{K,2} & w^{[1]}_{K,3} & \dots  & w^{[1]}_{K,20}
\end{bmatrix}
\]

According to Figure \ref{fig:simplified net}, the whole process to work on input $x$ can be summarized by Formula (\ref{equ:network}), where $x$ is a matrix containing from $x[n-K]$ to $x[n]$ and its dimension is $(1*5)$. 
\begin{equation}
\begin{split}
&o^{[1]} = xW^{[1]} + b^{[1]}\\
&a^{[1]} = a(o^{1})\\
&o^{[2]} = a^{[1]}W^{[2]} + b^{[2]}\\
&a^{[2]} = a(o^{2})\\
&o^{[3]} = a^{[2]}W^{[3]} + b^{[3]}\\
&y_{e}[n] = o^{[3]}
\end{split}
\label{equ:network}
\end{equation} 
  
\paragraph{Loss Function}
We choose the Mean Square Error (MSE) as the loss function, because the difference between an estimate $y_{e}[n]$ and the real $y[n]$ could be an either positive or negative number. The optimization goal of this DNN model is to generate the $y_{e}[n]$ as similar as possible to $y[n]$. As a result, the optimization during the DNN training is formulated as:
\begin{equation}
loss = \displaystyle\sum\limits_{m=0}^{M}||y_{real}[m]-y_{e}[m]||^2
\label{equ:loss}
\end{equation}
where $M$ is the mini-batch size of training data and $y_{e}[m]$ is an output of the DNN, which estimates the SI channel output, and $y_{real}[m]$ is a collected real SI sample coming  through the SI channel. Each parameter of the DNN network such as $W^{[l]}$ and $b^{[l]}$ will be updated through back-propagation in every epoch by deviating this loss function.

\subsubsection{DNN Training}
To train this DNN, we need to construct a training dataset to be in the dimension of $((N-K+1)*K)$ due to the input $x$ dimension is $(1*K)$, where $N$ is total number of samples in training dataset. To achieve this, a recurrent scheme is employed to generate $K$ samples for each input. For example, suppose the input samples are $[s_{1},s_{2},s_{3},s_{4},s_{5},s_{6},s_{7},s_{8},s_{9}]$, and $K$ equal to 3, the dataset is constructed as below :

\vspace{-0.2in}
\[ dataset = 
\begin{bmatrix}
    s_{1} & s_{2} & s_{3} \\
    s_{2} & s_{3} & s_{4} \\
    s_{3} & s_{4} & s_{5} \\
    \vdots & \vdots & \vdots\\
    s_{7} & s_{8} & s_{9} \\
\end{bmatrix}
\]
   
Then a training mini-batch is generated by randomly selecting a mini-batch size number of rows from the dataset. The benefit of using mini-batch to feed data is the faster and more stable convergence of loss function compared to using one sample at one time, because the loss derivation takes more than one samples so that any exceptional sample will only have little impact on the direction of convergence.

\subsection{Depp Learning Digital Cancellation Framework}\label{sec:steps}
\begin{figure}[h]
\vspace{-0.1in}
\includegraphics[width=.5\textwidth]{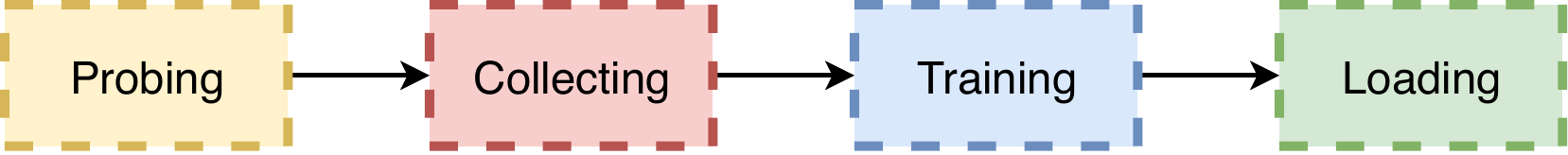}
\caption{Basic Steps of Our Design}
\vspace{-0.2in}
\label{fig:steps}
\end{figure}

With the SI channel modeled by a DNN discussed above, our proposed dee learning digital cancellation framework, as illustrated in \ref{fig:steps}, includes: (1) SI channel probing, (2) SI channel data collection, (3) DNN SI channel model training, and (4) DNN model data loading for digital cancellation, which are presented in details in the following.

\subsubsection{SI Channel Probing}
The first step of our design is to probe the SI channel. In this period, an IBFD wireless node sends designated probing signals or vectors $x$ from its transmission chain while its receiver chain senses what is received. The designated probing signals will pass the SI channel and transform to unknown received SI signals denoted by $y_{si}$.

\subsubsection{SI Channel Data Collection}
Following the SI channel probing, the framework collects and records the $I$ and $Q$ components of both the probing complex signals $x$ and the received SI signals $y_{si}$. Both of them are baseband complex signals, and their $I$ and $Q$ components comprise constellation points on a modulation constellation map. 

\subsubsection{DNN SI Channel Model Training}
With the data of $x$ and $y_{si}$ collected, they are then used to train a DNN SI channel model in a supervised learning way. The structure of the DNN SI channel model is explained in details in the next section \ref{sec:net}.

\subsubsection{Loading for Digital Cancellation}
After the DNN SI channel model is sufficiently trained, the trained model is loaded into the receiver chain. In IBFD wireless communication later, this trained DNN channel model estimates the SI signal $\hat{y}_{si}$ according to the Formula (\ref{equ:network}). Meanwhile, the receive chain recovers the received useful signal $m$ based on Formula (\ref{equ:eva_y_si}) with the SI signal $\hat{y}_{si}$ estimated by the trained DNN SI channel model.

\section{ Prototype Implementation}\label{sec:experiments}
To perform the assessment in real world, we have implemented our solution into a SDR prototype testbed of GNURadio and two USRP X310 nodes. 

\subsection{Prototype Architecture}
\begin{figure}[h]
\vspace{-0.2in}
\centering
\includegraphics[width=.4\textwidth]{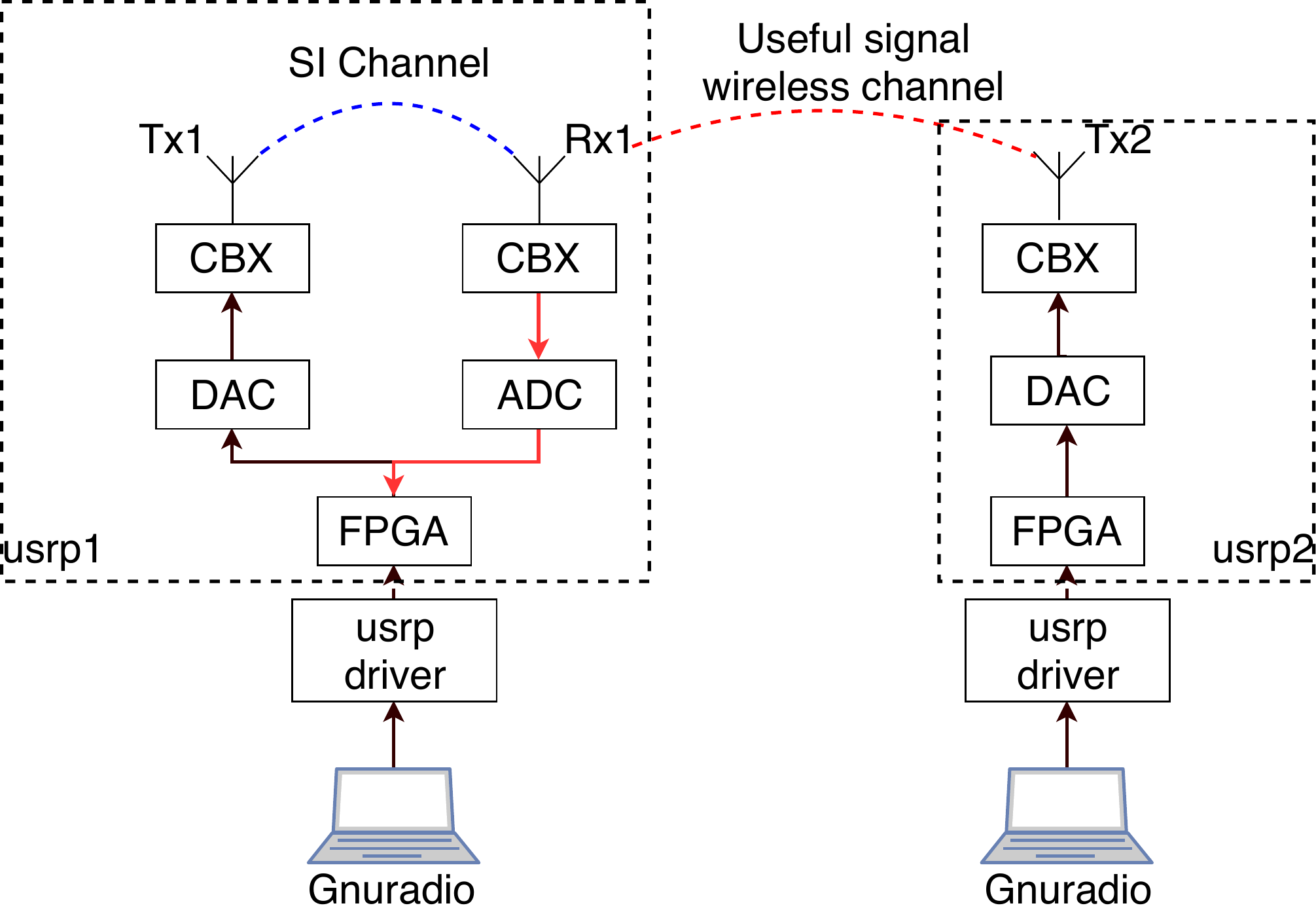}
\caption{Testbed Prototype Architecture}\label{fig:structure}
\vspace{-0.15in}
\end{figure}

The architecture of our prototype is illustrated in Figure \ref{fig:structure} showing how GNURadio and USRP operate together. GNURadio is the software that can customize the wireless radio and works with UHD, the USRP node driver. USRP X310 nodes are generic SDR hardware. To implement our solution to work in the real world, several challenges are confronted.

\subsection{Modulation in Gnuradio}
The first challenge is in modulation. To achieve a higher data transmission rate, modulation functions should be implemented. The Modulation block in GNURadio takes byte stream as input, and output $I-Q$ complex samples. If we want to apply our design on those $I-Q$ complex samples, we have to design a way transforming message data from byte stream to those complex samples.
\begin{figure}[h]
\vspace{-0.1in}
\centering
\includegraphics[width=.5\textwidth]{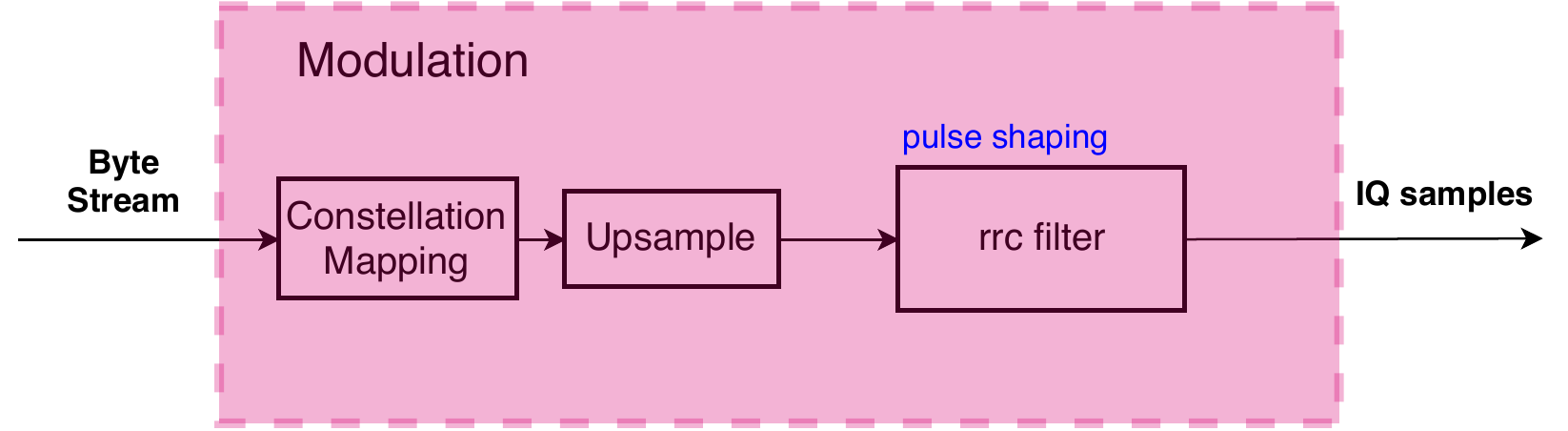}
\caption{Inner Block in Modulation}\label{fig:gnuradio_mod}
\vspace{-0.1in}
\end{figure}

\subsubsection{Constellation Mapping}
\begin{figure}
	\vspace{-0.3in}
    \centering
    \begin{subfigure}[b]{0.2\textwidth}
        \includegraphics[width=\textwidth]{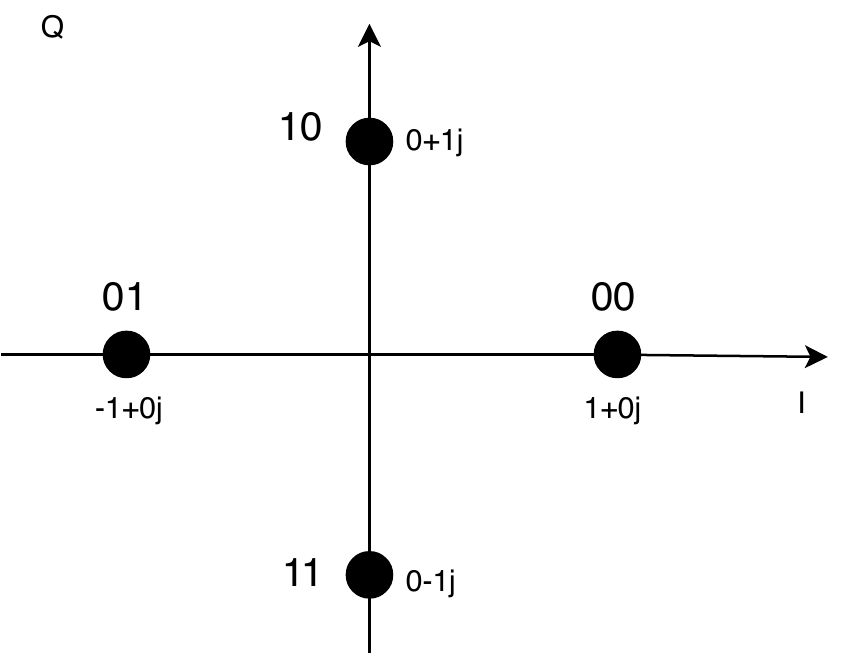}
        \caption{Constellation Mapping}
        \label{fig:constel}
    \end{subfigure}
    ~ 
    \begin{subfigure}[b]{0.2\textwidth}
        \includegraphics[width=\textwidth]{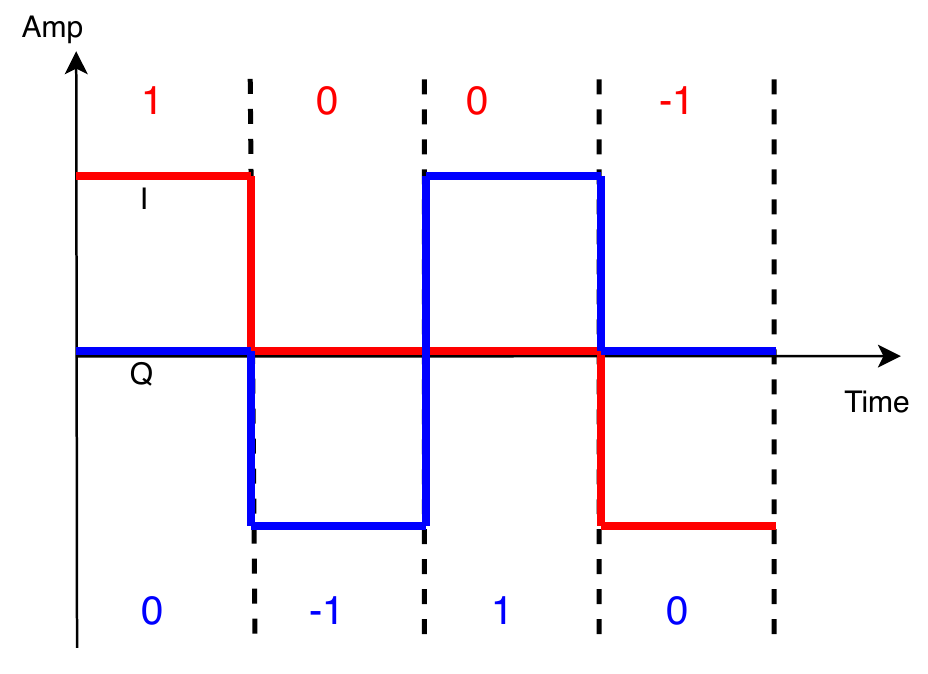}
        \caption{I/Q pulse}
        \label{fig:pulse}
    \end{subfigure}
    ~ 
    \begin{subfigure}[b]{0.2\textwidth}
        \includegraphics[width=\textwidth]{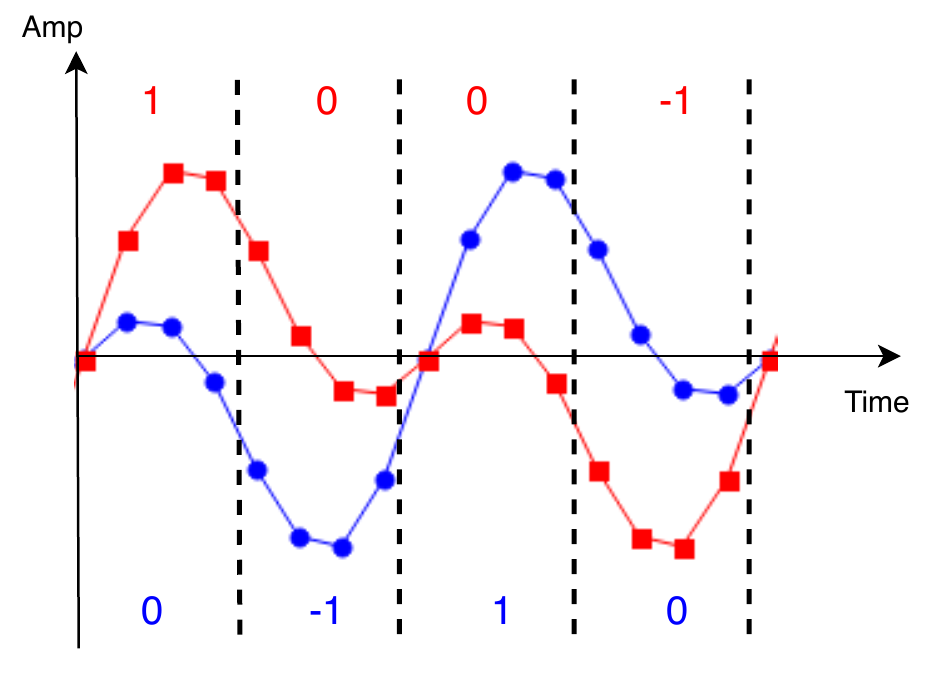}
        \caption{Pulse shaping}
        \label{fig:pulse_shaping}
    \end{subfigure}
    \caption{Modulation Process}\label{fig:mod}

\end{figure}

In GNURadio, the first block in Modulation is the Constellation Mapping block, which takes $k$ bits, where $k = log_2{M}$, and $M$ is number of constellation points, and generates output pulse of In-phase (I) and Quadrature (Q) signals. Suppose the input is a char $57$, which takes one byte in a bit sequence of $00111001$. Take QPSK as example, in this case, $M=4$ and $k=2$, if the constellation map is designed as Figure (\ref{fig:constel}). Every time only two bits of the bit sequence $00111001$ are modulated, i.e. $00$  to $1+0j$, $11$ to $0-1j$, $10$ to $0+1j$ and $01$  to $-1+0j$. As a result, the modulated complex sequence is $[1+0j,0-1j,0+1j,-1+0j]$ for this bit sequence as shown in Figure (\ref{fig:pulse}).

\subsubsection{Up-sampling and Pulse shaping}
After symbols are generated as shown in Figure (\ref{fig:pulse}),  they are processed with up-sampling. The up-sampling ratio depends on a sample-per-symbol parameter. For example, an  up-sampling ratio of 4 means every symbol is represented by 4 samples. Noted that a sample in digital informtion is a sampling point in a continuous signal. Since square pulse has large ISI, a pulse shaping filter follows up-sampling, which is a root-raised-cosine filter to reshape the pulse to transmit in the channel. This filter results in a better frequency domain response, and it achieves $0$ ISI in theory when two root-raised-cosine filters are combined together as one raised cosine filter as in Equation (\ref{equ: rrc filter}). Figure (\ref{fig:pulse_shaping}) shows what I/Q samples look like after up-sampling and pulse shaping. Those samples are then the exact inputs to the SI channel, and are what we need to collect.
\begin{equation}
H_{rc}(f) = H_{rrc}(f)\cdot H_{rrc}(f)
\label{equ: rrc filter}
\end{equation}

\subsection{Synchronized Data Collection}
The second challenge to the prototype is how to synchronize the collected channel data. It is very difficult to collect synchronized input signals to SI channel and their corresponding output signal from the SI channel because there is a random delay in the signal processing chain of USRP hardware. Collecting the input signals to the SI channel data starts as soon as the program starts. However, the output signal from the SI channel data to be collected by the receiving chain of the USRP node is randomly delayed. This delay can be host and USRP communication delay or driver communication delay.

As for the USRP random delay problem, researchers have presented a method to address this issue by calculating the barker code correlation \cite{suksmono2013simple}. A barker code is added at the head of every source signal sequence. At a receiver, once if the receive chain has all barker codes in its buffer, it can extract a synchronized signal. We have implemented a bark code based frame synchronization block tin GNURadio, to identify a synchronized signal in the frame synchronization block by inserting a probe function in main stream of GNURadio signal processing chain.

\subsection{Loading Module}
Another implementation challenge is about loading the DNN model for digital cancellation, because there is no deep learning module in GNURadio. After we collect synchronized SI channel input and output data and train the DNN channel model, we save all of the DNN parameters such as $W^{[l]}$ and $b^{[l]}$ into a dictionary data structure. The DNN model is implemented in Pytorch, but a loading module block is created in GNURadio to load the trained model parameter dictionary, and the known input signal is forwarded to this model to generate an estimated $y_e[n]$. Algorithm {\bf 1} below shows the implementation of loading and SI cancellation.
\vspace{-0.1in}
\begin{algorithm}[h]
\label{algo}
\caption{SI Cancellation} 
\hspace*{0.02in} {\bf Input:} 
input vector source $x[n]$, baseband complex signal $y(t)$\\
\hspace*{0.02in} {\bf Output:} 
output vector source $\hat{m}[n]$\\
\hspace*{0.02in} {\bf Description:} 
This algorithm input $x[n]$ and $y(t)$, where $y(t)$ is sum complex signals contains SI signal $\hat{x}(t)$ and useful signal $\hat{m}(t)$, then cancel $\hat{x}_{si}(t)$ from $y(t)$ to recover $\hat{m}[n]$
\begin{algorithmic}[1]
\State $N = log_2 M$
\For{$k < length(x[n])$} 
 　　\State $x(t) = $Modulate(x[k], N)
 	\State $k=k+1$
\EndFor

\State $\hat{x}_{si}(t) = load\_model([x(t),x(t-1),...,x(t-taps+1)],path)$
\State $\hat{m}(t) = y(t)-\hat{x}_{si}(t)$
\State $\hat{m}[n] = Demodulate(\hat{m}(t), N)$
\State \Return $\hat{m}[n]$
\vspace{-0.05in}
\end{algorithmic}
\end{algorithm}

\section{Performance}\label{sec:performance}
We have conducted extensive performance evaluation of our solution with the implemented prototype in real world.

\subsection{DNN Loss Convergence}
\begin{figure}[h]
	\vspace{-0.15in}
    \centering
    \begin{subfigure}[b]{0.2\textwidth}
        \includegraphics[width=\textwidth]{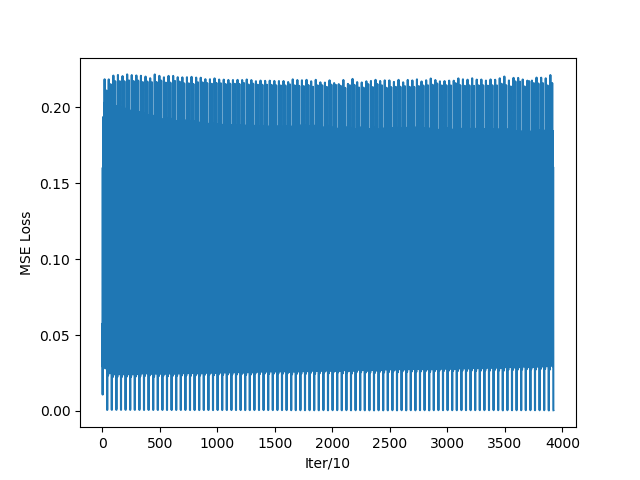}
        \caption{Sample Size $K$= 2}
        \label{fig:loss2}
    \end{subfigure}
    ~ 
    \begin{subfigure}[b]{0.2\textwidth}
        \includegraphics[width=\textwidth]{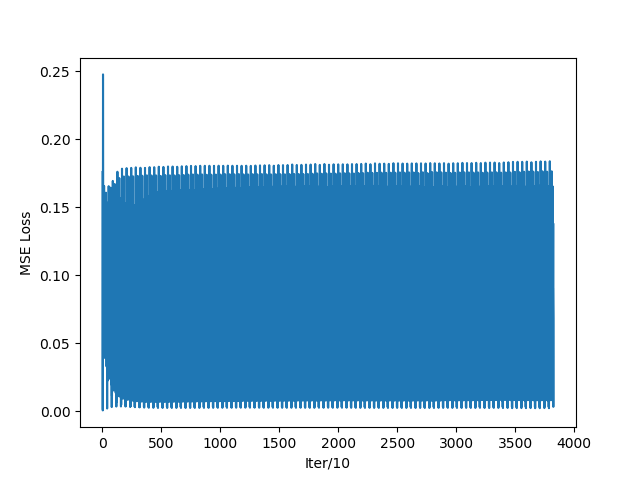}
        \caption{Sample Size $K$= 20}
        \label{fig:loss20}
    \end{subfigure}
    ~ 
    \begin{subfigure}[b]{0.2\textwidth}
        \includegraphics[width=\textwidth]{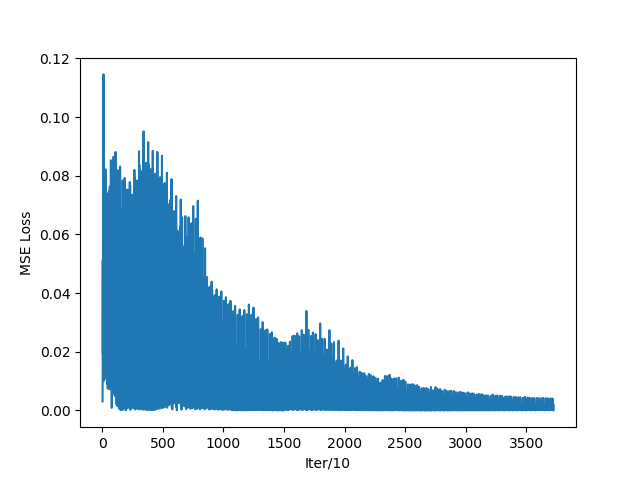}
        \caption{Sample Size $K$= 50}
        \label{fig:loss50}
    \end{subfigure}
    \begin{subfigure}[b]{0.2\textwidth}
        \includegraphics[width=\textwidth]{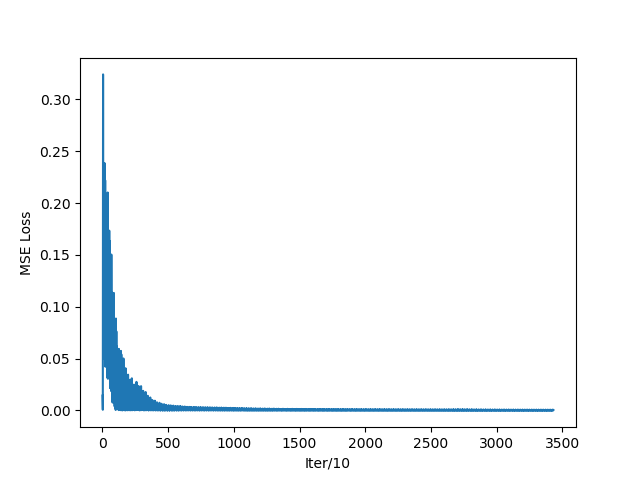}
        \caption{Sample Size $K$= 100}
        \label{fig:loss100}
    \end{subfigure}
    \caption{Loss Convergence at Various Input Sample Sizes}\label{fig:loss}
    \vspace{-0.2in}
\end{figure}

First, we have tested various DNNs with different $K$ to see the impact of the input size on the DNN parameter convergence. Figure \ref{fig:loss} shows the convergence dynamics of the MSE loss with $K$=2, 20, 50 and 100 respectively. Because each byte contains $8$ bits, every $2$ bits are modulated into a QPSK symbol, and each symbol is represented by $4$ samples, $16$ samples consequently are needed to represent one byte. Therefore, any sample size smaller than 16 does not contribute to the learning. This is why the sample size $2$ gives nothing for the network to learn as illustrated on the figure. When the sample size is larger than $16$, the DNN learns something as shown in Figure \ref{fig:loss20}. With the taps size increased to $100$, the loss convergences very quick and stays at a very low level, which means the DNN model can generate a predicted SI $y_{si}[n]$ very close to the real SI $y_{real}[n]$.

\subsection{SI Estimation and Cancellation}
\begin{figure}[h]
\centering
\includegraphics[width=.5\textwidth]{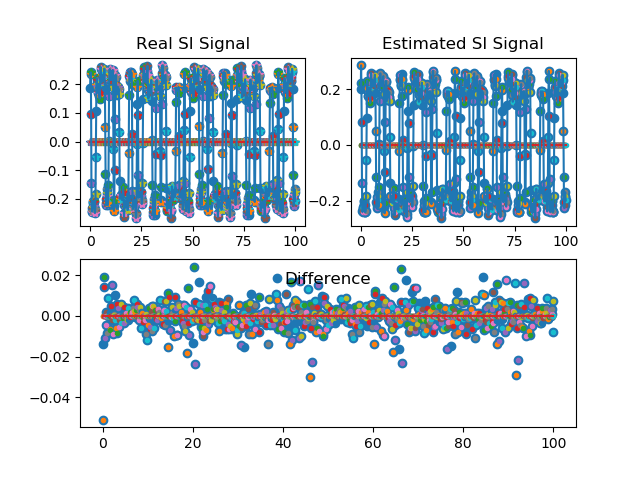}
\vspace{-0.2in}
\caption{SI Estimation with 100 Taps}
\label{fig:taps100}
\vspace{-0.2in}
\end{figure}

We have then tested the performance of the DNN channel model in estimating the SI after the training. Figure \ref{fig:taps100} shows the difference between the real SI signals and the estimated SI signals. The difference is smaller than 0.02, which means this solution can cancel at least $10\log_{10} \frac{0.3}{0.02}=7dB$, and the average performance is $10\log_{10} \frac{0.3}{0.005}=17.8dB$.

\subsection{Bit Error Rate with SI Cancellation}
Further experiments have been conducted to assess the impact of various modulations on the bit error rate in four environment settings: Room1, Room2, Outdoor, and Hallway. We transmit a vector stream $[1,2,3,...,10]$ in transmission chain with three types of modulations (QPSK, 16PSK, and 64PSK) at the USRP node1, then cancel the SI in GNURadio with the trained model, next demodulate the complex signals after the SI cancellation to get the vector stream $[101,102,103,...,120]$ transmitted by the USRP node2. After that, the reconstructed vector stream after the SI cancellation is finally compared with the original vector stream sent by the USRP2 node2 to calculate the bit error rates. These experiments have been repeated for many times in each of the four scenarios, and the average bit error rates are calculated as shown in Table \ref{table:psk}.
\begin{table}[h]
\vspace{-0.1in}
\centering
\begin{tabular}{|l|c|c|c|}
\hline
Scenario/BER(\%)   & QPSK & 16PSK & 64PSK\\
\hline
Room1		  & 8.5 & 25 & 37.4\\ \hline
Room2		  &	11.2 & 28  & 41.6\\\hline
Outdoor		  & 8.0 & 21 & 37.9 \\	\hline	
Hallway		  & 13.3 & 27.5 & 45.1 \\	\hline
\end{tabular}
\caption{BER vs Modulations}
\label{table:psk}
\vspace{-0.15in}
\end{table}

From the data in the table, the outdoor environment yields the best performance with our solution. The worst result is in the hallway, which indicates multi-path fading has a large impact on the performance. Another observation is that the performance degrades as the modulation level increases, which matches to the expectations that lower modulation level is more robust to transmission errors.

\section{Conclusion}\label{sec: conclusion}
In this work, we propose a novel non-linear digital cancellation method based on deep learning to achieve IBFD wireless. This solution first probes the channel to collect SI channel data, which are used to train the SI channel DNN. The trained parameters are used to estimate the SI for cancellation in the realtime communication. We have implemented this deep learning SI cancellation solution into a GNURadio prototype testbed with USRP nodes. The extensive performance evaluation shows that (1) the deep learning approach is feasible to achieve very high accuracy in digital cancellation, (2) the proposed deep learning model has a fast and stable convergence in the training, (3) the deep learning solution can work across various modulations in different environments. As other signal processing modules such as error coding are added, it is reasonable to expect better performance.

\section{Acknowledgement}
This material is based upon work supported by the National Science Foundation under Grant No. \#1726017.

\bibliographystyle{unsrt}
\bibliography{def.bib}
\end{document}